\documentclass[%
 reprint,
 amsmath,amssymb,
 aps,
]{revtex4-2}

\usepackage[english]{babel}
\usepackage{graphicx,import}
\usepackage{dcolumn}
\usepackage{bm}

\usepackage[utf8]{inputenc}
\usepackage[T1]{fontenc}
\usepackage{mathptmx}
\usepackage{etoolbox}

\usepackage{tikz}
\usetikzlibrary{matrix}
\usetikzlibrary{fit}
\usetikzlibrary{backgrounds}
\usetikzlibrary{positioning}
\usetikzlibrary{arrows}
\usetikzlibrary{shapes}
\usetikzlibrary{calc}
\usepackage{circuitikz}
\usepackage{setspace}
\usepackage{multirow}
\usepackage{adjustbox}
\usetikzlibrary{decorations.pathreplacing, angles, quotes}

\usetikzlibrary{arrows.meta, positioning, chains}
\usepackage{tikz}
\usetikzlibrary{matrix}
\usetikzlibrary{fit}
\usetikzlibrary{backgrounds}
\usetikzlibrary{positioning}
\usetikzlibrary{arrows}
\usetikzlibrary{shapes}
\usetikzlibrary{calc}
\usepackage{circuitikz}
\usepackage{setspace}

\tikzstyle{state}=[rectangle,
    thick,
    minimum height=0.6cm,
    minimum width=1.0cm,
    rounded corners]

\tikzstyle{state1}=[state,
    draw=orange!80,
    fill=orange!20]

\tikzstyle{state2}=[state,
    draw=teal!80,
    fill=teal!20]

\tikzstyle{state3}=[state,
    draw=blue!80,
    fill=blue!20]

\tikzstyle{state4}=[state,
    draw=brown!80,
    fill=brown!20]

\tikzstyle{state5}=[state,
    draw=yellow!80,
    fill=yellow!20]

\tikzstyle{line} = [draw, -latex']

\tikzstyle{auto}=[rectangle,
    thick,
    minimum height=0.6cm,
    minimum width=0.6cm,
    draw=purple!80,
    fill=purple!20,
    rounded corners]

\tikzstyle{background}=[rectangle,
    fill=gray!10,
    inner sep=0.15cm,
    rounded corners=3mm]

\begin{document}

\preprint{APS/123-QED}

\title{Attention-Enhanced Reservoir Computing}

\author{Felix Köster}
 \email{felixk@mail.saitama-u.ac.jp}
\author{Kazutaka Kanno}%
\author{Jun Ohkubo}%
\author{Atsushi Uchida}%
 \email{auchida@mail.saitama-u.ac.jp}
\affiliation{%
 Department of Information and Computer Sciences, Saitama University 255 Shimo-Okubo, Sakura-ku, Saitama City, Saitama, 338–8570, Japan
}%

\date{\today}

\begin{abstract}
Photonic reservoir computing has been successfully utilized in time-series prediction as the need for hardware implementations has increased. Prediction of chaotic time series remains a significant challenge, an area where the conventional reservoir computing framework encounters limitations of prediction accuracy. We introduce an attention mechanism to the reservoir computing model in the output stage. This attention layer is designed to prioritize distinct features and temporal sequences, thereby substantially enhancing the prediction accuracy. Our results show that a photonic reservoir computer enhanced with the attention mechanism exhibits improved prediction capabilities for smaller reservoirs. These advancements highlight the transformative possibilities of reservoir computing for practical applications where accurate prediction of chaotic time series is crucial.
\end{abstract}

\maketitle

\section{\label{Sec1}Introduction}

The combination of photonics and machine learning has emerged as a promising field of information processing, utilizing the development of more rapid and efficient computational methodologies \cite{shen2017deep}. Central to this interdisciplinary area is the concept of photonic reservoir computing, a paradigm that utilizes optical capabilities to enhance reservoir computations \cite{larger2012photonic, Nakajima2021, Brunner2013ParallelPI, Nakayama:16, Bueno:17, 8765366, Takano:18, Sackesyn21}. A semiconductor laser with optical feedback is one example of a photonic reservoir, which is a prime example of the photonic delay-based reservoir computer model \cite{Brunner2013ParallelPI, Nakayama:16, Bueno:17, 8765366, Takano:18}. Numerous studies have underscored the efficacy of delay-based reservoir computing in tackling a variety of complex tasks, extending from bit error correction in optical communications \cite{Sackesyn21} to prediction of spatiotemporal chaotic systems \cite{Rafayelyan2020}.

Chaotic time-series prediction is known to be challenging owing to its inherent sensitivity to accuracy and its pronounced nonlinearity \cite{strogatz:2000}. A prime example of this challenge is weather forecasting, which is typically limited to short-term predictions due to the fundamentally chaotic nature of weather systems.

The framework of reservoir computing has been widely implemented using a variety of physical devices, such as electronic circuits \cite{appeltant2011information}, spintronic devices \cite{torrejon2017neuromorphic}, soft robots \cite{ef0a6b769141456283377bec586ba791}, and nanostructure materials \cite{kotooka2021}.
One of the advantages of reservoir computing is the easy implementation of the reservoir without high computational cost because input weights and network weights are randomly fixed, and only the output weights are trained by a learning algorithm.
Classical reservoir computing is often limited because of the inflexibility of the fixed weight structure when faced with the complexities inherent in such chaotic time series \cite{jaeger2001echo}. However, hardware technology advancements have opened pathways for accelerated prediction, giving access to real-time responses and improved computational efficiency \cite{tanaka2019recent}. This progress has enabled the integration of reservoir computing with other advanced machine learning methods, one example being deep learning and its most successful recent development, the attention mechanism \cite{vaswani2017attention}.

To further enhance the predictive capabilities of reservoir computing, we propose integrating the attention mechanism within a reservoir computer. This mechanism has demonstrated superior effects in diverse areas of machine learning \cite{vaswani2017attention}. By incorporating an attention layer as the output layer, the system acquires the capacity to prioritize specific temporal characteristics, potentially yielding a more dynamical comprehension of the data \cite{bai2018empirical}.

This study introduces the attention mechanism to delay-based photonic reservoir computing using a semiconductor laser with optical feedback. 
We perform chaotic time series prediction by deploying two benchmark tasks. 
The first is the unidirectionally-coupled two-Lorenz systems, in which the reservoir computer is only exposed to the $x$, $y$, and $z$ variables of the forced Lorenz system \cite{lorenz1963deterministic}, while the driving system remains hidden, thus rendering the task a non-Markovian deterministic chaotic system with hidden states.
The second is a time series consisting of alternating Lorenz and Rössler data sets \cite{Kanno2020AdaptiveMS}, which we use to investigate how well the attention mechanism adapts to swiftly changing tasks.

\section{Attention-Enhanced Reservoir Computing}
\label{sec:att_enh_res}

\subsection{Reservoir Computing}

Reservoir Computing (RC) is a paradigm in the field of recurrent neural networks that offers an efficient approach to tasks involving temporal dynamics and sequential patterns \cite{tanaka2019recent}. 
The delay-based RC is a prime example of RC, distinctive for its utilization of time-multiplexed masking for input representation \cite{appeltant2011information}. 

Let us assume we have a data set $\mathbf{X}=\{(x_1, y_1),\dots(x_l, y_l),\dots, (x_L, y_L) \}$ where we want to approximate $y_l$ through $x_l$, where $l$ is the indexing and $L$ the number of data points in the set.
We multiply every data point \( x_l\) with a periodic mask $m(t)$, which is essentially a high-dimensional random vector distributed over time. This mask expands and distributes the data point \( x_l\) to enrich the temporal response dynamics of the reservoir system. We therefore have the masked input \( v_l(t) \), formally represented as \( v_l(t) = x_l \cdot m(t) \), where \( m(t) \) is the masking vector of length $T$. 
In this study, we use a mask $m(t)$ that is constant for an interval, where the values of the mask $m_n$ are drawn randomly from a uniform distribution on the interval between 0 and 1. 
$N$ is the number of reservoir node states within $T$, i.e., $T=N\theta$.
After applying $v_l(t)$ to the reservoir, we sample the system's response at distances of $\theta$ yielding a high-dimensional representation, allowing for rich and diverse dynamics to be captured for downstream tasks.

As a  reservoir, we use a delay-based photonic reservoir computer, which captures the dynamics of semiconductor lasers with weak optical feedback. The Lang-Kobayashi equations mathematically represent the systems as follows \cite{Brunner2013ParallelPI, Nakayama:16, Uchida1994OpticalCW}.
\begin{align}
\dot{E}(t) &= (1 + i \tilde{\alpha}) E(t) - |E(t)|^2 E(t) + \kappa E(t - \tau) \exp(i \phi) \\
\dot{n}(t) &= p + \eta v_l(t) - (1 + |E(t)|^2)n(t) \label{eq2}
\end{align}
where, \( E(t) \) denotes the complex electric field amplitude, \( n(t) \) represents the carrier density, \( \tilde{\alpha} \) is the linewidth enhancement factor, \( \kappa \) signifies the feedback strength, \( \tau \) is the feedback delay time, \( \phi \) is the phase shift, and \( p \) is the normalized injection current. 
The parameter values are summarized in Table \ref{table:parameters}.

Within the context of delay-based RC, the input modulates the injection current $p$, thus affecting the system's dynamics.
The modulated injection current is denoted as $\eta v_l(t)$, where $\eta$ is the input strength and $v_l(t)$ is the input to the reservoir system. The reservoir responses \( r(t) \) are derived from the intensity \(r(t) = I(t) = |E(t)|^2 \) of the electric field amplitude at discrete time points of distance $\theta$, i.e. $r(\theta), r(2\theta), \dots, r(N\theta)$.
To yield the response of one input data point $x_l$, we have
\begin{align}
 \textbf{r}_l = \begin{pmatrix} r(lT + \theta),r(lT + 2\theta) &\dots, & r(lT + N\theta)
\end{pmatrix}^T,
\end{align}
with $^T$ denoting the transpose of the vector $\textbf{r}_l$.
Collecting all the responses for all the data points $x_l$, we arrive at the response matrix \( \textbf{R} \in \mathbb{R}^{L \times N} \) with 
\begin{align}
\textbf{R} = \begin{pmatrix} \textbf{r}_1^T \\
\vdots \\ 
\textbf{r}_L^T
\end{pmatrix}
\end{align}

In the classic RC framework, only the readout weights are trained while the internal weights of the reservoir are left unchanged \cite{jaeger2001echo}.
The output or readout weights are crucial in mapping the reservoirs states to the desired outputs. Denoted typically as $\textbf{w}_{\text{reg}} \in \mathbb{R}^{N \times 1}$, these weights are applied to the reservoir states $\textbf{R}$ to approximate the target $\textbf{y} \in \mathbb{R}^{L \times 1}$. 
Training the output weights involves optimizing to minimize the distance between the approximation vector $\textbf{d}=\textbf{R}\textbf{w}_{\text{reg}} \in \mathbb{R}^{L \times 1}$ and the true target vector $\textbf{y}$. 
Commonly, a linear regression approach is employed \cite{jaeger2001echo}, considering a regularization term to prevent overfitting. The training process can be mathematically represented as:
\begin{equation}
\textbf{w}_{\text{reg}} = \arg \min_{\textbf{w}} \left(|| \textbf{y} - \textbf{R} \textbf{w} ||^2 + \lambda ||\textbf{w}||^2\right)
\label{eq:ridge_target}
\end{equation}
where \( \textbf{R} \) denotes the collection of reservoir states, \( \textbf{y} \) represents the targets, and \( \lambda \) is the regularization coefficient.

The analytical solution to the ridge regression problem can be derived by setting the gradient of Eq. \eqref{eq:ridge_target} with respect to \( \textbf{w} \) to zero. The resulting equation can be expressed as:
\begin{equation}
\textbf{w}_{\text{reg}} = (\textbf{R}^T \textbf{R} + \lambda \textbf{I})^{-1} \textbf{R}^T y,
\end{equation}
where \( I \in \mathbb{R}^{N \times N}\) denotes the identity matrix. This solution provides a direct way to compute the optimal \( \textbf{w}_{\text{reg}} \) without iterative optimization. The addition of the regularization term \( \lambda \textbf{I} \) ensures that the matrix \( \textbf{R}^T \textbf{R} + \lambda \textbf{I} \) is invertible, thus guaranteeing a unique solution and penalizing overfitting.

\begin{table}[ht]
\centering
\caption{Summary of parameters and their numerical values.}
\begin{tabular}{|m{5cm}|m{3cm}|}
\hline
\textbf{Parameter} & \textbf{Numerical Value} \\ \hline
Mask period \( T \) & $N\theta = 5 \times 10^{-9}$ s\\ \hline
Linewidth enhancement factor \( \tilde{\alpha} \) & $3$ \\ \hline
Phase shift \( \phi \) & $0.0$ \\ \hline
Feedback strength \( \kappa \) & $10^{8}$ \\ \hline
Feedback delay time \( \tau \) & $1.01T = 5.05 \times  10^{-9}$ \\ \hline
Normalized injection current \( p \) & $1.11$ \\ \hline
Input strength \( \eta \) & $0.08$ \\ \hline
Number of nodes \( N \) & $50$ \\ \hline
Interval length \( \theta \) & $10^{-10}$ s \\ \hline
\end{tabular}
\label{table:parameters}
\end{table}

\subsection{Attention Mechanism}

In deep learning, particularly when dealing with sequential data, the attention mechanism has emerged as a groundbreaking concept \cite{vaswani2017attention}. This mechanism enables models to selectively focus on relevant portions of the input sequence during output generation, mirroring how humans selectively prioritize information during decision making or data processing. Central to this mechanism is the calculation of attention weights, which subsequently guide a weighted combination of input values. These weights effectively signify the level of "attention" or importance each part of the input should receive.

To clarify the method of attention, we present an example inspired by Ref. \cite{zhang2023dive}. Imagine a database containing daily temperature records over several years, illustrated as tuples such as \{(``2024-01-01'', ``-2°C''), (``2024-01-02'', ``0°C''), (``2024-01-03'', ``3°C'')\}, where the dates act as keys ($k$) and temperatures as values ($v$). Querying this database ($q$) with the date ``2024-01-02'', would fetch the exact temperature ``0°C''. When an exact match for the queried date is absent and approximate matching is allowed, the system might return approximate temperatures close to neighboring dates.
In this example, the query ($q$) asks for the attention considering all values in the sequence (here being three temperatures at different time steps).
The keys ($k$) enable to focus on specific values ($v$) of the input sequence by comparing the query ($q$) to all the keys ($k$) of the input sequence.

Expanding on this illustration, we compute a distance function between the query ($q$) and the keys ($k$) within the entire input sequence $\mathbf{U}=\{(k_1, v_1),\dots, (k_m, v_m) \}$, providing a weighting (or attention) to the associated values ($v$). This relationship is captured as:
\begin{align}
    \text{attention}(\textbf{q},\textbf{K}) = \sum_{i=1}^m \alpha(\textbf{q},\textbf{k}_i)\textbf{v}_i,
    \label{eq:att}
\end{align}
where $\text{attention}(\textbf{q},\textbf{K})$ is the weighted sum of all values in the input sequence for a given query ($\textbf{q}$), $i$ indexes the input sequence $\mathbf{U}$, and $\alpha(\textbf{q},\textbf{k}_i)$ denotes the scalar attention weights. An interesting aspect is the determination of the function $\alpha$. Several functions can be employed, from traditional distances like the Boxcar function, defined as $\alpha_{\text{Boxcar}} = 1$ \quad if \quad $||\textbf{q} - \textbf{k}_i|| \leq 1$, to a Gaussian similarity, given by $\alpha_{\text{Gaussian}} = \exp(||\textbf{q} - \textbf{k}_i||^2_2)$ \cite{zhang2023dive}. In contemporary deep learning, the function $\alpha$ is often approximated using neural networks, allowing gradient descent to tailor the function optimally to the task itself.

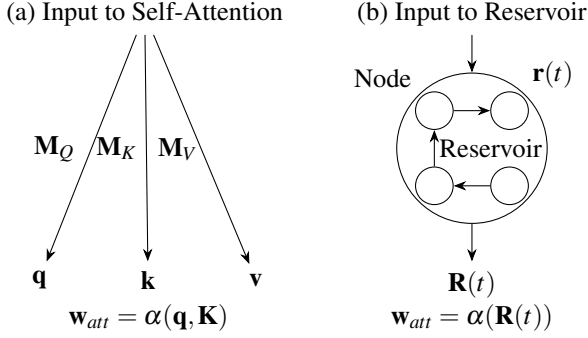
\begin{figure}
    \centering
  \begin{tikzpicture}[>=Stealth, node distance=1cm]
    \node[circle, draw, minimum size=2cm] (reservoir) {};

    \node[circle, draw, minimum size=0.5cm] at ([xshift=5mm, yshift=5mm]reservoir.center) (inner1) {};
    \node[circle, draw, minimum size=0.5cm] at ([xshift=-5mm, yshift=-5mm]reservoir.center) (inner2) {};
    \node[circle, draw, minimum size=0.5cm] at ([xshift=-5mm, yshift=5mm]reservoir.center) (inner3) {};
    \node[circle, draw, minimum size=0.5cm] at ([xshift=5mm, yshift=-5mm]reservoir.center) (inner4) {};

    \draw[->] (inner2) -- (inner3);
    \draw[->] (inner3) -- (inner1);
    \draw[->] (inner4) -- (inner2);

    \node at (-4.35,1.79) (data1) {(a) Input to Self-Attention};
    \node[above=0.5cm of reservoir] (data2) {(b) Input to Reservoir};

    \node at (-5.75,-1.75) (q1) {$\textbf{q}$}; 
    \node[right=of q1] (k1) {$\textbf{k}$};
    \node[right=of k1] (v1) {$\textbf{v}$};

    \draw[->] (data1) -- (q1) node[midway, left, xshift=-1mm] {$\textbf{M}_Q$};
    \draw[->] (data1) -- (k1) node[midway, left] {$\textbf{M}_K$};
    \draw[->] (data1) -- (v1) node[midway, right, xshift=-7mm] {$\textbf{M}_V$};

    \node at (-4.3,-2.25) {$\textbf{w}_{att} = \alpha(\textbf{q},\textbf{K})$};

    \node at (0,-2.25) {$\textbf{w}_{att} = \alpha(\textbf{R}(t))$};
    
    \node at (0.25,0) {Reservoir};

    \draw[->] (data2) -- (reservoir) node[midway, left, xshift=-3mm] {};
    \draw[->] (reservoir.south) -- ++(0,-0.5) node[below] {$\textbf{R}(t)$};

    \node[above left=0mm of inner3] {Node};
    \node[above right=0mm of inner1] {$\textbf{r}(t)$};

\end{tikzpicture}
    \caption{(a) Self-attention mechanism creating queries ($\textbf{q}$), keys ($\textbf{k}$), and values ($\textbf{v}$) from an input. (b) Reservoir approach transforming an input into a high dimensional reservoir state $\textbf{R}(t)$. Both projections are then used in an abstract function $\alpha$ to yield the attention weights $\textbf{w}_{att}$.}
    \label{fig:self-att_res}
\end{figure}

A more recent progression in attention mechanisms introduces the concept of self-attention \cite{bahdanau2014neural, vaswani2017attention}. In this framework, the queries ($\textbf{q}$), keys ($\textbf{k}$), and values ($\textbf{v}$) originate from the same data point as shown in Fig. \ref{fig:self-att_res}(a). For each individual data point in the input sequence (e.g, the temperature at a specific time step), corresponding queries ($\textbf{q}$), keys ($\textbf{k}$), and values ($\textbf{v}$) are produced. Typically, these elements are generated through matrix transformations, where the matrix weights are refined and optimized using a gradient descent algorithm.

\subsection{Incorporation of Attention Layer in Reservoir Computing}

Next, we delve into incorporating the attention mechanism within a reservoir computer.
While numerous potential locations for integrating the attention mechanism exist, we focus on its application to improve the reservoir computer's output layer. Therefore we replace the traditional weights $\textbf{w}_{\text{reg}}$ with the attention weights $\textbf{w}_{\text{att}}$.

\begin{figure}%
	\centering
   	\includegraphics[width=0.5\textwidth]{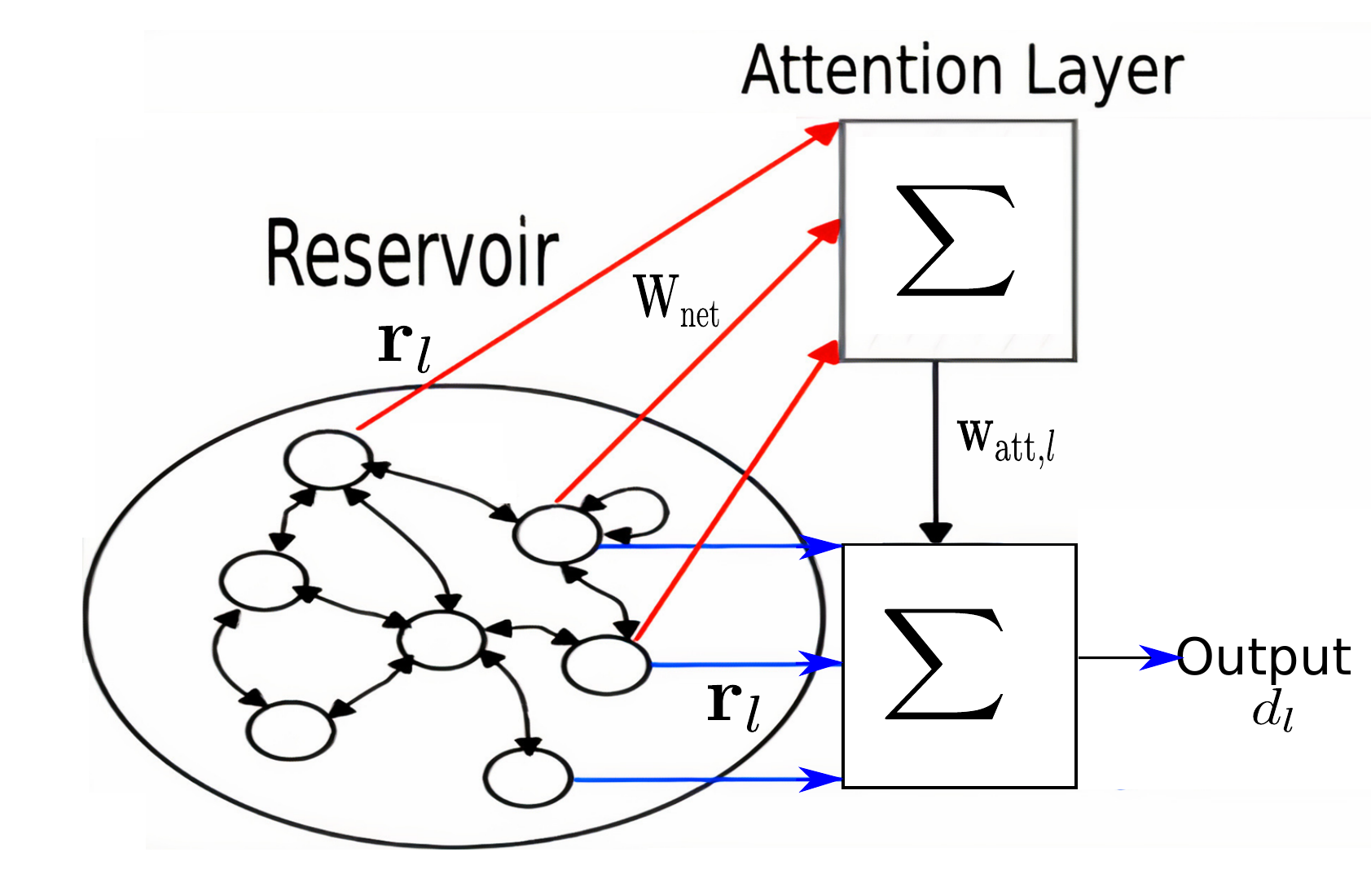}
	\caption[]{Schematic of the attention-enhanced reservoir computing setup. The trainable parameters $\textbf{W}_{\text{net}}$ take the reservoir node states and yield the attention weights $\textbf{w}_{\text{att}}$, which are then weighted with the reservoir node states to generate an output. }
	\label{fig:sketch_liner_attention}
\end{figure}%

In our attention-enhanced reservoir computer, the reservoir states $r(t)$ simultaneously serve as the queries, keys and values, as shown in Fig. \ref{fig:self-att_res}(b). This is reminiscent of the self-attention scheme where every data point of the input sequence is projected onto queries, keys, and values, often via matrices learned through a gradient descent algorithm \cite{bahdanau2014neural, vaswani2017attention}. 
In the case of reservoir computing, the reservoir projects the input into a high-dimensional feature space, similar to the matrix transformation that projects the input data point to the queries, keys, and values.

The concept of attention-enhanced reservoir computing is depicted in Fig. \ref{fig:sketch_liner_attention}. We consider one-dimensional target signals.
In the self-attention mechanism, the query of one data point in the sequence is combined with the keys of all other data points in the sequence to generate the attention weights, which are then weighted with the values (see \cite{bahdanau2014neural, vaswani2017attention}).
In the attention-enhanced scheme, the role of the reservoir node states $\textbf{r}_l \in \mathbb{R}^{N \times 1}$ is equal to the queries, keys, and values of the self-attention approach, where $N$ is the number of reservoir nodes and $l \in [1,L]$ is the $l$-th data point applied to the reservoir.
To predict the $l$-th data point, we concisely write the attention-enhanced reservoir computer via:
\begin{align}
    \textbf{w}_{\text{att},l} &= \textbf{W}_{\text{net}} \textbf{r}_l, \label{eq7} \\
    d_l &= \textbf{w}_{\text{att},l}^T \textbf{r}_l, \label{eq8}
\end{align}
where $d_l$ represents the approximation value for a target signal $y_l$, $\textbf{w}_{\text{att},l} \in \mathbb{R}^{N \times 1}$ are the input-dependent attention weights, and $\textbf{W}_{\text{net}} \in \mathbb{R}^{N \times N}$ are the attention-layer weights that are trained via a gradient descent algorithm. 
Here, Eq. \eqref{eq7} represents the attention layer, taking the reservoir node state $\textbf{r}_l$ as input and yielding the attention weights $\textbf{w}_{\text{att},l}$.
Eq. \eqref{eq8} represents the output layer of the reservoir computer, computing the weighted sum of the reservoir nodes to approximate the target $y_l$, in which the classic ridge regression weights are now substituted by the attention weights $\textbf{w}_{\text{att}}$.
A detailed description of a multidimensional target signal is given in Appendix \ref{sec:prediction_inference}.

We point out that the scheme shown in Eqs. \eqref{eq7} and \eqref{eq8} is linear and a substitution of Eq. \eqref{eq7} into Eq. \eqref{eq8} is possible.
However, we still keep the equations separated into Eq. \eqref{eq7} and Eq. \eqref{eq8} as it depicts the general architecture of the attention-enhanced approach.
For example, we show a result of a nonlinear attention approach in Appendix \ref{app:non_att}.

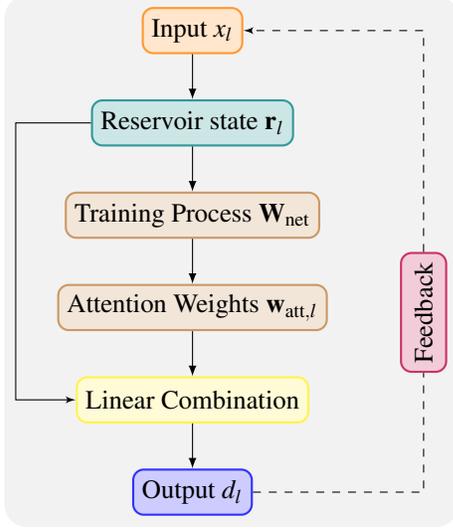
\begin{figure}[h]
\centering
\begin{tikzpicture}[>=latex]
  \matrix (mtrx) [row sep=0.6cm, column sep=0.2cm, matrix of nodes, nodes in empty cells] {
    & \node (standardized_data) [state1]{Input \(x_l\)}; & \\
    & \node (stand_reservoir_state) [state2]{Reservoir state \(\textbf{r}_l\)}; & \\
    & \node (nn) [state4]{Training Process \(\textbf{W}_{\text{net}}\)}; & \\
    & \node (weights) [state4]{Attention Weights \(\textbf{w}_{\text{att},l}\)}; & \\
    & \node (linear_comb) [state5]{Linear Combination}; & \\
    & \node (prediction) [state3]{Output \(d_l\)}; & \\
    };
    
    \path[->]
    (standardized_data) edge node[pos=0.5,right] {}  (stand_reservoir_state)
    (stand_reservoir_state) edge node[pos=0.5,right] {} (nn)
    (nn) edge node[pos=0.5,right] {} (weights)
    (weights) edge node[pos=0.5,right] {} (linear_comb)
    (linear_comb) edge node[pos=0.5,right] {} (prediction);
    
    \node (auto) [auto, right=of mtrx-4-1, xshift=3.75cm, rotate=90, anchor=north] {Feedback};
    
    \path [line,dashed] (prediction.east) -- ++(2.27,0) -- (auto.west)
        (auto.east) -- ++(0,+1.5) |- (standardized_data.east);
        
    \path [line] (stand_reservoir_state.west) -- ++(-1,0) coordinate (leftPoint) |- (linear_comb.west);
    
    \begin{pgfonlayer}{background}
        \node [background,
                    fit=(standardized_data) (prediction) (auto) (leftPoint)] {};
    \end{pgfonlayer}
\end{tikzpicture}
\caption{Schematic representation of the reservoir computing process augmented with an attention mechanism.}
\label{tikz_diagram}
\end{figure}

\subsection{Procedure of Attention-Enhanced Reservoir Computing}

A detailed visualization of our complete scheme's data stream is shown in Fig. \ref{tikz_diagram}. We describe this representation by providing a step-by-step explanation of the underlying operations as follows.

\subsubsection{Reservoir Processing}
The standardized data, \( x_l \), is passed into the reservoir, resulting in a reservoir state. In our setup, \( x_l \) is processed in Eq. \eqref{eq2} via the injection current of the semiconductor laser. This reservoir state is an intermediate, high-dimensional representation of the input data. To standardize the data, we modify the training data set to have the mean of zero and standard deviation of one for training and testing.


\subsubsection{Training Process}
The standardized reservoir state, \( \textbf{r}_l \), undergoes processing through a linear transformation with weights \( \textbf{W}_{\text{net}} \in \mathbb{R}^{N \times N} \). 
Regarding the training of the weights \( \textbf{W}_{\text{net}} \), deep learning methods can be applied~\cite{goodfellow2016deep} using a gradient descent algorithm:
\begin{align}
    \textbf{W}_{\text{net}}(s+1) = \textbf{W}_{\text{net}}(s) -\gamma \nabla F(\textbf{W}_{\text{net}}(s), \textbf{R}),
\end{align}
with $\gamma$ being a learning rate, which we set to $0.01$, and $s$ iterates through the gradient descent epochs, and the optimization objective is:
\begin{align}
    \textbf{W}_{\text{net}} = \min_{W} \left(F(\textbf{W}, \textbf{R}, \textbf{y})\right),
\end{align}
where $\textbf{R}$ is the set of all reservoir states and $\textbf{y}$ is the target dataset.
We deploy $F$ as a simple squared distance, i.e. $F(\textbf{W}, \textbf{R}, \textbf{y}) = \frac{1}{2} ||d(\textbf{R},\textbf{W}) - \textbf{y}||_2^2$, with $d(\textbf{R},\textbf{W})$ being the approximation to the target dataset $\textbf{y}$.
To clarify, the loss $F(\textbf{W}, \textbf{R}, \textbf{y})$ (in our case the NRMSE) of the full training data set of all $K$ data points is calculated for every epoch and the weights $\textbf{W}_{\text{net}}$ are optimized via the gradient descent.
To speed up the process, first we calculate the reservoir states for all $K$ data points and save them.
After that we use the reservoir states to calculate $F(\textbf{W}, \textbf{R}, \textbf{y})$ and apply the gradient descent.

The attention-enhanced reservoir thus loses its advantage of utilizing the analytically solvable ridge regression as an optimization approach.
Owing to the recent improvement in hardware-accelerated optimization of gradient descent approaches, we believe that the limiting computation factor of reservoir computing is not the output layer optimization. Therefore, augmenting reservoir computing with the attention mechanism could couple the best of two methods.
The loss over the epochs of the gradient descent is depicted in the Appendix \ref{sec:loss}.

\subsubsection{Attention Weights}
The attention weights,  \( \textbf{w}_{\text{att},l} \), emerge from the linear transformation \( \textbf{W}_{\text{net}} \). These weights dictate how much importance is placed on each node or neuron in the reservoir at a specific time. In essence, the attention mechanism dynamically adjusts, prioritizing certain nodes over others based on their relevance to the output at that particular time step. 
The attention weights \( \textbf{w}_{\text{att},l} \) are computed via a linear transformation of the standardized reservoir state \( \textbf{r}_l \) via the weights \( \textbf{W}_{\text{net}} \), represented as:
\begin{align}
    \textbf{w}_{\text{att},l} = \textbf{W}_{\text{net}}\textbf{r}_l
\end{align}

It is worth noting that the weights $\textbf{W}_{\text{net}}$ are static after training. The time dependence of the attention weights $\textbf{w}_{\text{att},l}$ arise from the fact that the reservoir node states are time-dependent. Therefore, a full photonic-based implementation of the attention-enhanced reservoir computer could be achieved.

\subsubsection{Linear Combination}
The attention weights and the standardized reservoir states are then linearly combined. This combination effectively amplifies the states that the attention mechanism deems significant. The corresponding output \( d_l \) is given by:
\begin{align}
    d_l = \textbf{w}_{\text{att},l}^T\textbf{r}_l
\end{align}

Our integration of the attention layer in reservoir computing enables the system to prioritize important features in the input data, particularly in complex time series prediction scenarios. The attention-enhanced reservoir computer offers a more nuanced and accurate representation of the data by dynamically shifting focus to the most important nodes in the reservoir based on context.

\subsection{Prediction Quality Metrics}
\label{sec:metrics}
In the context of reservoir computing for time series prediction, two primary configurations are commonly used: the open-loop and closed-loop \cite{cucchi:hal-03774230}. 

\subsubsection{Normalized Root-Mean-Square Error}

In the open-loop configuration, the model performs a one-step-ahead predictions. We evaluate the model’s prediction accuracy via the Normalized Root-Mean-Square Error (NRMSE) \cite{james2013introduction} as: 
\begin{align}
    \text{NRMSE} = \sqrt{\frac{\sum_{l=1}^{L} || d_l - y_l||^2}{N \text{Var}(Y)}},
\end{align}
where $y_l$ is the $l$-th target data point, $d_l$ is the prediction data, $l$ is the index up to the $L$-th data point, and $\text{Var}(Y)$ is the variance over the whole data set.

\subsubsection{Valid Prediction Time}

Fig. \ref{tikz_diagram} illustrates the closed-loop feedback configuration. In this setup, after the model has been trained, it is switched to a closed-loop mode, where the model's own predictions are recursively fed back as input. 
This method allows for continuous predictions, extending beyond the one-step-ahead prediction. The quality of predictions in a closed-loop configuration is evaluated by a metric called the valid prediction time (VPT) \cite{PhysRevLett.120.024102}, defined as:
\begin{align}
    \text{VPT} = \delta_y(t) > 0.4, \quad \delta_u(t) = \frac{|y(t) - d(t)|^2}{\langle|y(t) - \langle y(t) \rangle |^2 \rangle},
\end{align}
where VPT measures the first time at which $\delta_y(t)$ surpasses the value of $0.4$, in which $\delta_y(t)$ measures the average normalized distance over the trajectory between the target $y(t)$ and the reservoirs output $d(t)$, and $\langle\rangle$ denotes the time average.
VPT measures the duration for which the model's predictions remain within an acceptable error threshold, thereby indicating the effectiveness of the model in maintaining reliable predictions over time.
Unlike open-loop predictions, closed-loop predictions compound potential errors over time, because of the sensitive dependence on initial conditions on chaotic dynamic systems.

\section{Numerical Results}
\label{Sec:results}

This section presents the comparative results of the attention-enhanced photonic reservoir computing to the conventional linear regression technique. The evaluation is performed for a unidirectionally-coupled two-Lorenz system introduced in Sec. \ref{sec:uni-coupled-lorenz} and an alternating Lorenz-Rössler system introduced in Sec. \ref{sec:RLS}.

\subsection{Unidirectionally-Coupled Two-Lorenz System (UCTLS)}
\label{sec:uni-coupled-lorenz}

\subsubsection{Model}

As a benchmark task, we utilize a unidirectionally-coupled two-Lorenz system (UCTLS)~\cite{lorenz1963deterministic, FEN2017200}, defined by the following equations:

\begin{align}
\label{eq:lorenz_lorenz_1}
  \dot{x}_1 &= (a + \sigma_{\text{force}} x_2) (y_1-x_1), \\
  \dot{y}_1 &= x_1(b + \sigma_{\text{force}} y_2 - z_1) - y_1, \\
  \dot{z}_1 &= x_1y_1 - (c + \sigma_{\text{force}} z_2) z_1, \\
   \dot{x}_2 &= a (y_2-x_2), \\
  \dot{y}_2 &= x_2(b - z_2) - y_2, \\
  \dot{z}_2 &= x_2 y_2 - c z_2.
  \label{eq:lorenz_lorenz_last}
\end{align}

These equations represent two Lorenz systems, with the first one being driven by the second. The parameters used, namely \(a=10\), \(b=28\), and \(c=\frac{8}{3}\), are used as the standard Lorenz parameters~\cite{lorenz1963deterministic}. The key distinction is the introduction of a forcing term, \(\sigma_{\text{force}}\), that couples the two systems unidirectionally.

For the reservoir computing task, only the $x_1$, $y_1$, and $z_1$ variables are exposed, while $x_2$, $y_2$, and $z_2$ are hidden.
Predicting such a non-markovian deterministic chaotic time series is challenging~\cite{abarbanel1993analysis}, especially when it has underlying hidden states that are not directly observable by the reservoir computer.
The system was integrated with a Runge-Kutta 45 approach with a step size of $dt=0.01$, while the sampling time is $dt_{\text{sample}}=0.1$.
Regarding the largest Lyapunov exponent of the system, our simulations indicate that no significant variation up to a force of $\sigma_{\text{force}}=0.15$ is exhibited relative to the classic Lorenz value of $\lambda_{\text{Lorenz}}\approx 0.91$.
For all simulations we use 25000 training and 5000 testing data points.

We begin with the UCTLS as defined by Eqs. \eqref{eq:lorenz_lorenz_1}-\eqref{eq:lorenz_lorenz_last} subjected to a perturbation strength of $\sigma_{\text{force}} = 0.05$, implemented in a photonic reservoir utilizing 50 nodes, with other parameters specified in Table 1.

\subsubsection{Results of Time Series and Weights}

\begin{figure}%
	\centering
   	\includegraphics[width=0.5\textwidth]{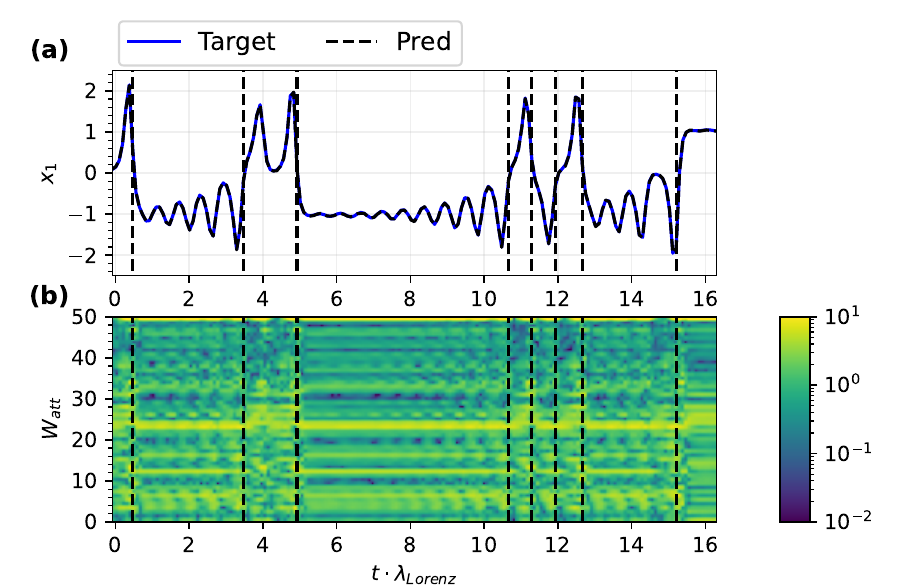}
	\caption[]{Example time series of open-loop configuration for the UCTLS. (a) shows the x-variable of the time series, and (b) the attention weights in the same time interval for a 50-node reservoir. The color bar shows the magnitude of the attention weights. The dashed black lines indicate a shift from the attractor, pinpointing the dynamical adjustment of the weights to the task.}
	\label{fig:Time_Series_Double_lorenz_Open_Loop}
\end{figure}%

Fig. \ref{fig:Time_Series_Double_lorenz_Open_Loop}(a) illustrates an open-loop time series. The upper graph displays the time series of the $x_1$-variable from the exposed Lorenz attractor of the true system (in blue) alongside the predicted time series (black line), plotted against the time in Lyapunov times with $\lambda_{\text{Lorenz}}\approx 0.91$. The similarity between the prediction and the actual system's behavior is accurate.

A more interesting insight comes from Fig. \ref{fig:Time_Series_Double_lorenz_Open_Loop}(b), which shows the computed absolute value of the attention weights $ \textbf{w}_{\text{att}}$ for each node within the reservoir across the same period. Nodes with higher weight magnitudes are assumed to be more important for predictions at specific time points. Vertical dashed lines indicate a pattern change corresponding to moments where the system's trajectory shifts from one side of the attractor to the other. A corresponding shift in the attention weights’ structure implies that the attention-enhanced reservoir can adapt dynamically to the time-dependent characteristics of the task. This feature is essential, as the UCTLS demands time-dependent transformations depending on the current position within the attractor space.

The attention mechanism provides a dynamic adjustment unlike the static nature of weights employed in linear regression, which cannot distinguish this characteristic. Given that nearly all complex systems evolve with time, adaptively adjusting the significance of different nodes is not just beneficial but necessary. This strongly illustrates the inherent limitations of static approaches when dealing with systems characterized by dynamic evolution.

\begin{figure}[t]
  \centering
    \centering
    \includegraphics[width=0.5\textwidth]{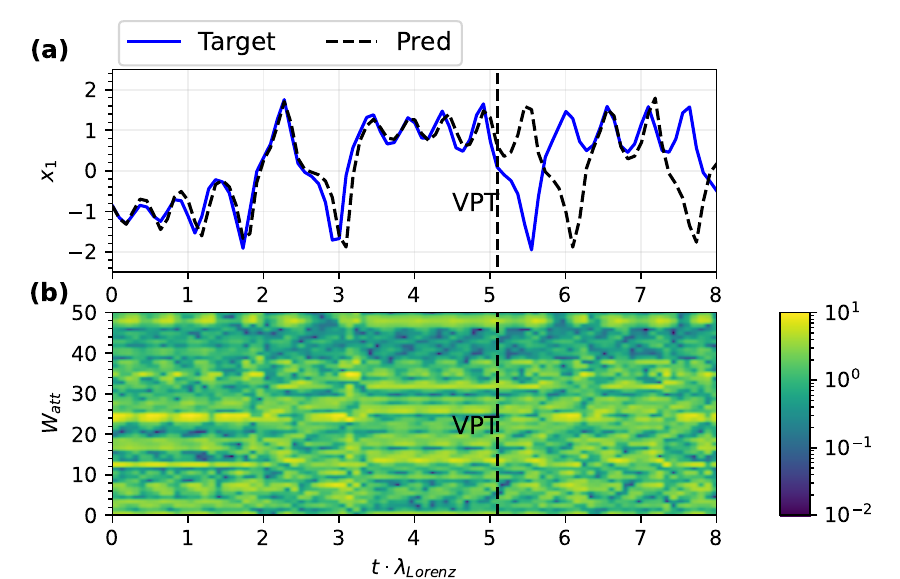}
    \caption[]{Example time series of closed-loop configuration for the UCTLS. (a) shows the x-variable of the time series, and (b) the attention weights in the same time interval for a 50-node reservoir. The true trajectory and closed-loop prediction are depicted by the solid blue and dashed black lines, respectively. The color bar shows the magnitude of the attention weights. The vertical dashed black line indicates the time point for VPT computation.}
    \label{fig:Time_Series_Double_lorenz}
\end{figure}

This behavior continues even if the closed-loop configuration is applied, as shown in Fig. \ref{fig:Time_Series_Double_lorenz}. The reservoir's prediction and the true trajectory stay close to each other until around 5 Lyapunov times, indicated by a vertical dashed black line pinpointing to the time at which the VPT (see Sec. \ref{sec:metrics}) is computed.
The weights in the lower graph show the distinctive feature of adjusting to the dynamical problem itself despite being configured in the closed-loop scheme.

\subsubsection{Results of Quality Metrics}

\begin{figure}%
	\centering
   	\includegraphics[width=0.5\textwidth]{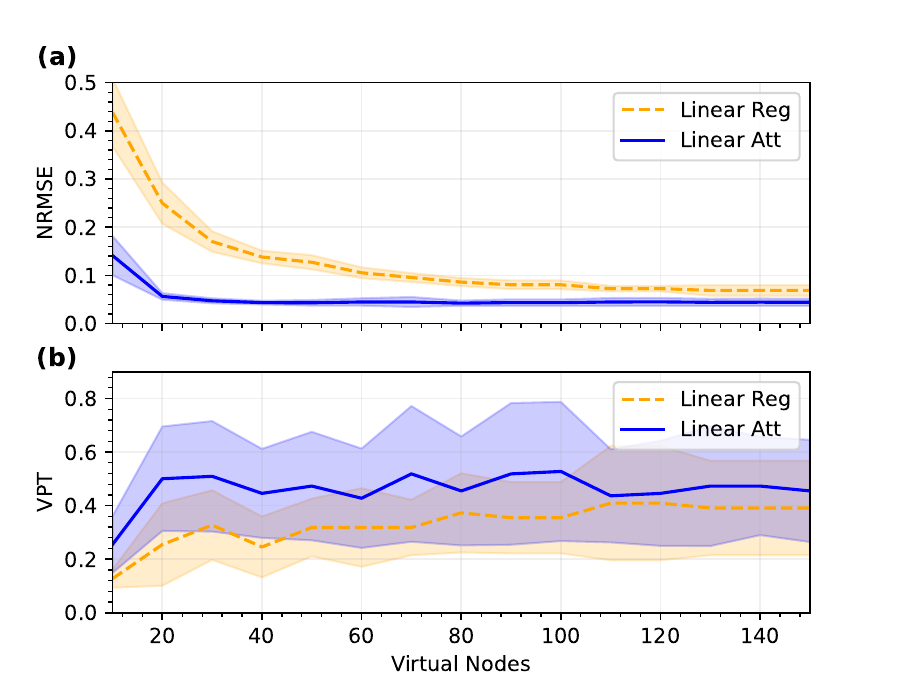}
	\caption[]{(a) NRMSE and (b) VPT for classical linear regression (dashed orange line) and attention-enhanced reservoir computing (solid blue line) with varying reservoir size $N$ for UCTLS. The shaded areas depict the standard deviations of NRMSE and VPT over ten time series.}
	\label{fig:NRMSE_VPT_res_size}
\end{figure}

To showcase the efficacy of dynamic adjustment, we simulated an ensemble prediction across ten distinct trajectories for varying reservoir sizes. The results are shown in Fig. \ref{fig:NRMSE_VPT_res_size}. We computed the NRMSE for the open-loop depicted in Fig. \ref{fig:NRMSE_VPT_res_size}(a) and the VPT for the closed-loop illustrated in Fig. \ref{fig:NRMSE_VPT_res_size}(b). These metrics were calculated for both the classical linear regression model (dashed orange line) and the attention-enhanced model (solid blue line) across a spectrum of reservoir sizes, from 10 to 150 nodes.
In all cases the same reservoir with the exact same reservoir states were used for both the classical approach and the attention-enhanced approach.

The results indicate that the attention-enhanced model yields a lower NRMSE for smaller-sized reservoirs while simultaneously achieving a higher VPT. Although this advantage diminishes as the reservoir size increases, it stays consistent. Notably, the optimal performance of the attention-enhanced reservoir is observed with as few as 20 nodes.
This suggests that the attention mechanism significantly reduces the required size of the reservoir. However, this size reduction comes at the costs of additional weights in the output layer. Nevertheless, this approach is particularly interesting when only limited reservoir sizes are realistically accessible.

\begin{figure}[t]
    \centering
    \includegraphics[width=0.5\textwidth]{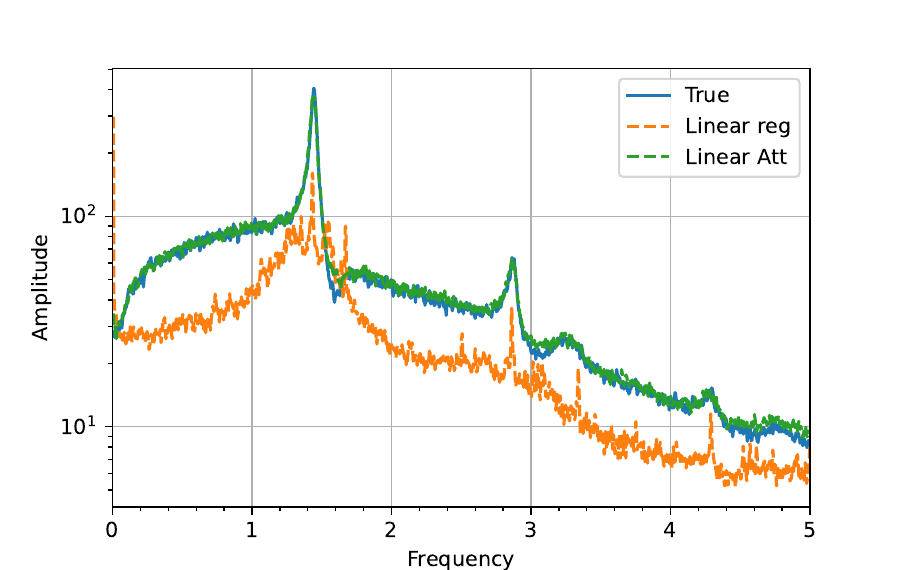}
    \caption[]{Fourier Spectrum of a predicted time series in the closed-loop configuration with a reservoir size of $N=30$. The true system is depicted as a solid blue line, while the classic linear regression and the attention-enhanced reservoir are shown as dashed orange and green lines, respectively}
    \label{fig:Fourier_Double_Lorenz}
\end{figure}

We analyze the average power spectrum of a time series obtained with a 30-node reservoir, reproducing the UCTLS task with \(\sigma_{\text{force}}=0.05\). Fig. \(\ref{fig:Fourier_Double_Lorenz}\) displays the resulting Fourier spectra: the true system's spectrum is depicted by the blue curve, the orange curve represents the closed-loop outcome of the classical linear regression approach, while the green curve corresponds to the linear attention approach.
An examination of Fig. \(\ref{fig:Fourier_Double_Lorenz}\) reveals that the linear regression approach significantly deviates in reconstructing the spectrum of the original attractor, failing to identify the correct main frequency and the secondary peak at a frequency of \(2.9\) Hz. In contrast, the linear attention method successfully captures both frequencies, reproducing the power spectrum very accurately.

\begin{figure}%
	\centering
   	\includegraphics[width=0.5\textwidth]{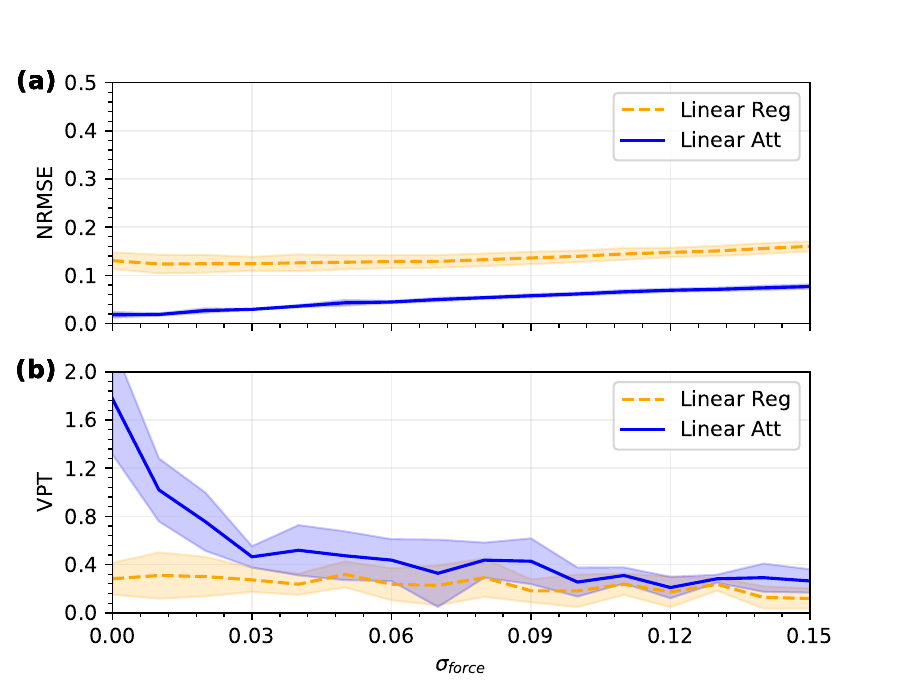}
	\caption[]{(a) NRMSE and (b) VPT for classical linear regression (dashed orange line) and attention-enhanced reservoir computing (solid blue line) with varying coupling strength $\sigma_{\text{force}}$ of UCTLS. The shaded areas depict the standard deviations of NRMSE and VPT over ten time series. The photonic reservoir size is set to $N=50$ nodes.}
	\label{fig:NRMSE_VPT_sigma_force}
\end{figure}

To deepen our analysis of the UCTLS, we evaluate the dependency of the NRMSE and VPT on the driving force \(\sigma_{\text{force}}\). We keep the reservoir size constant with $N=50$ nodes, while the other parameters are unchanged. The results are illustrated in Fig.
\(\ref{fig:NRMSE_VPT_sigma_force}\), using the same color coding and line style as in Fig. \(\ref{fig:NRMSE_VPT_res_size}\). The behavior of the classic single Lorenz system is represented on the left side of the graph with a driving force of \(\sigma_{\text{force}}=0\). The linear attention method outperforms the classical linear regression approach for the VPT and NRMSE metrics in this regime.
When the forcing strength is increased, the observed improvement for VPT diminishes, while the NRMSE gradually increases for both methods. However, the attention-enhanced approach yields lower NRMSE values throughout. These results underline the improved prediction performance of the attention-enhanced reservoir relative to the traditional approach.

\subsection{Alternating Lorenz and Rössler System (ALRS)}
\label{sec:RLS}

\subsubsection{Model}

The second benchmark is a target dataset constructed using two canonical chaotic systems: the Rössler and Lorenz models \cite{Kanno2020AdaptiveMS}. For the Rössler model, the equations of motion are given by:
\begin{align}
\frac{dx_R}{dt} &= -y_R - z_R, \label{eq:lorenz_roessler_1} \\
\frac{dy_R}{dt} &= x_R + a_Ry_R, \\
\frac{dz_R}{dt} &= b_R + x_Rz_R - c_Rz_R.
\end{align}
For the Lorenz model, the dynamics are described by:
\begin{align}
\frac{dx_L}{dt} &= a_L(y_L - x_L), \\
\frac{dy_L}{dt} &= -x_Lz_L + b_Lx_L - y_L, \\
\frac{dz_L}{dt} &= x_Ly_L - c_Lz_L.
\label{eq:lorenz_roessler_last}
\end{align}
The parameters used in our simulations are \(a_R = 0.2, b_R = 0.2, c_R = -5.7\) for the Rössler system and \(a_L = 10, b_L = 28, c_L = -\frac{8}{3}\) for the Lorenz system. Both of these systems are known for generating deterministic chaos. For data generation, we exclusively consider the \(x\)-variable trajectories (\(x_R\) for Rössler and \(x_L\) for Lorenz) if the open-loop configuration is applied, while for the closed-loop configuration the \(x\), \(y\), \(z\) variables are exposed. Therefore, a multidimensional target is used for attention/enhanced reservoir computing, as described in Appendix \ref{sec:prediction_inference}. 

The Rössler system is integrated with a time step of $dt=0.05$, while every ten points are downsampled in an effective time step of $dt_{\text{sample}}=0.5$. Similarly, the Lorenz system is integrated using a time step of $dt=0.01$ and subsequently downsampled to achieve an effective time step of $dt_{\text{sample}}=0.1$. For each system, multiple 500-step trajectories are generated. These trajectories are then alternatively concatenated, yielding a time series that alternates between Rössler and Lorenz every 500 points for 25000 training data points and 5000 testing data points \cite{Kanno2020AdaptiveMS}. Before concatenation, the time series data from both models are standardized to ensure zero mean and unit standard deviation.
We call this benchmark the ALRS.
For both systems, the largest Lyapunov exponents are calculated with $\lambda_{\text{Lorenz}} \approx 0.91$ and $\lambda_{\text{Rössler}} \approx 0.071$.

\subsubsection{Results}
\begin{figure}%
	\centering
   	\includegraphics[width=0.5\textwidth]{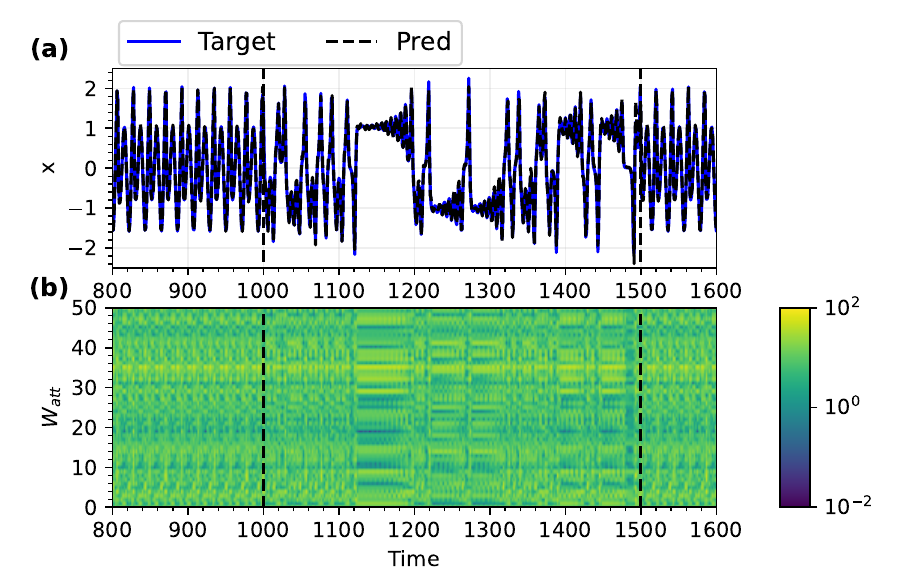}
	\caption[]{Example time series of open-loop configuration for ALRS. (a) shows the x-variable of the time series, and (b) the attention weights in the same time interval. The color bar shows the magnitude of the attention weights. The dashed black lines indicate the change in task from Rössler to Lorenz or vice versa, pinpointing to the dynamical adjustment of the weights to the task.}
	\label{fig:Time_Series_Lorenz_Roessler}
\end{figure}%

We investigate the ALRS, defined by Eqs. \(\eqref{eq:lorenz_roessler_1}\) to \(\eqref{eq:lorenz_roessler_last}\) in the same manner. An exemplary time series of this system in the open-loop configuration is depicted in Fig. \(\ref{fig:Time_Series_Lorenz_Roessler}\). Fig. \(\ref{fig:Time_Series_Lorenz_Roessler}\)(a) indicates that the prediction closely approximates the true system, suggesting a high degree of accuracy in predicting the next state.

Fig. \(\ref{fig:Time_Series_Lorenz_Roessler}\)(b) provides the interesting part of the attention-enhanced reservoir. The vertical dashed lines denote the transitions between the Lorenz and Rössler dynamics within the task. A shift in the attention weights in response to these changes can be observed, implying that the attention-enhanced reservoir can adapt to different dynamical problems presented by the input signal.

\begin{figure}%
	\centering
   	\includegraphics[width=0.5\textwidth]{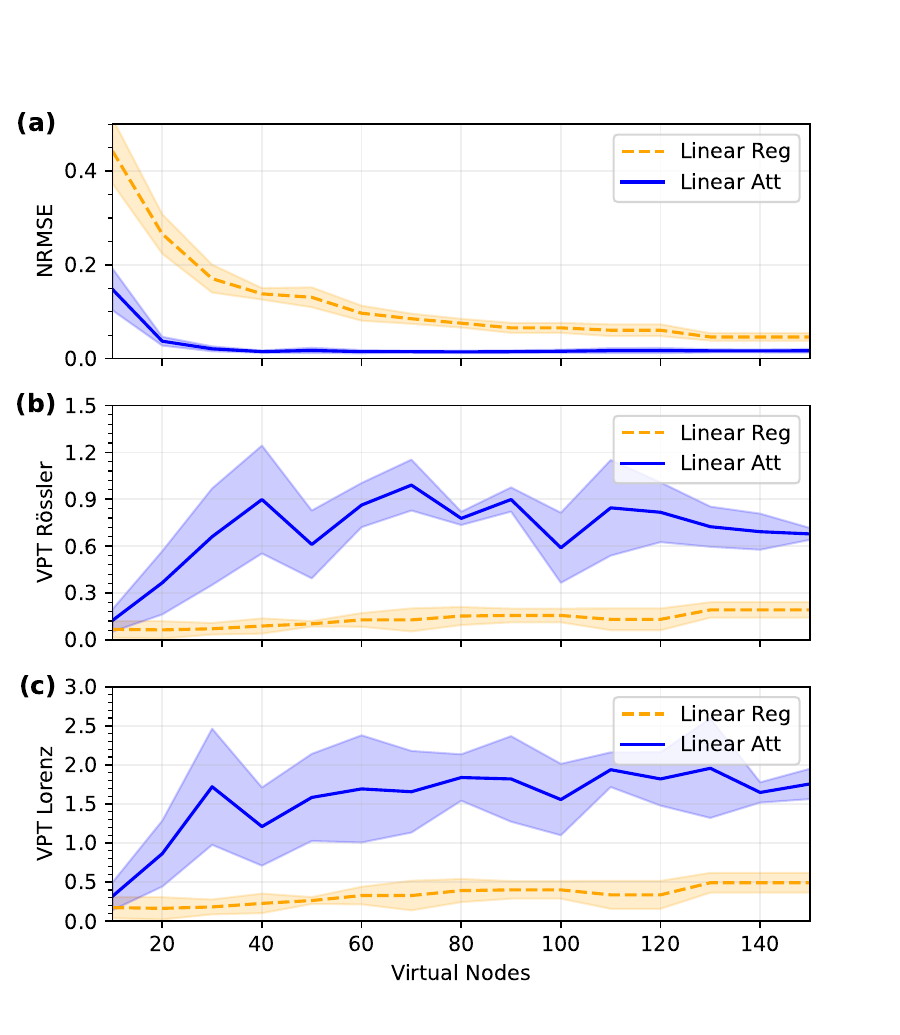}
	\caption[]{
 (a) NRMSE, (b) VPT for UCTLS, and (c) VPT for ALRS for classical linear regression (dashed orange line) and attention-enhanced reservoir computing (solid blue line) with varying coupling strength reservoir sizes $N$. The shaded areas depict the standard deviations of NRMSE and VPT over ten time series.}
	\label{fig:NRMSE_VPTS_Lorenz_Roessler}
\end{figure}

To quantitatively analyze the prediction quality, we computed the NRMSE for the open-loop and the VPT for the close- loop in ALRS across reservoir sizes, spanning from 10 to 150 nodes, as shown in Fig. \(\ref{fig:NRMSE_VPTS_Lorenz_Roessler}\). 
Note that in the ALRS benchmark, two different VPTs are computed by starting the closed-loop configuration either in a window of Rössler inputs or in a window of Lorenz inputs.
Here, the NRMSE for the classical linear regression method is presented by the dashed orange line, while the solid blue line represents the attention approach. These findings agree with previous observations: the attention-enhanced reservoir yields smaller NRMSE (higher VPT), with the effect being particularly pronounced for smaller reservoir sizes.
Interestingly, this dynamic adjustment can tackle a problem that swiftly changes between two distinct attractors.

\begin{figure}[t]%
	\centering
   	\includegraphics[width=0.5\textwidth]{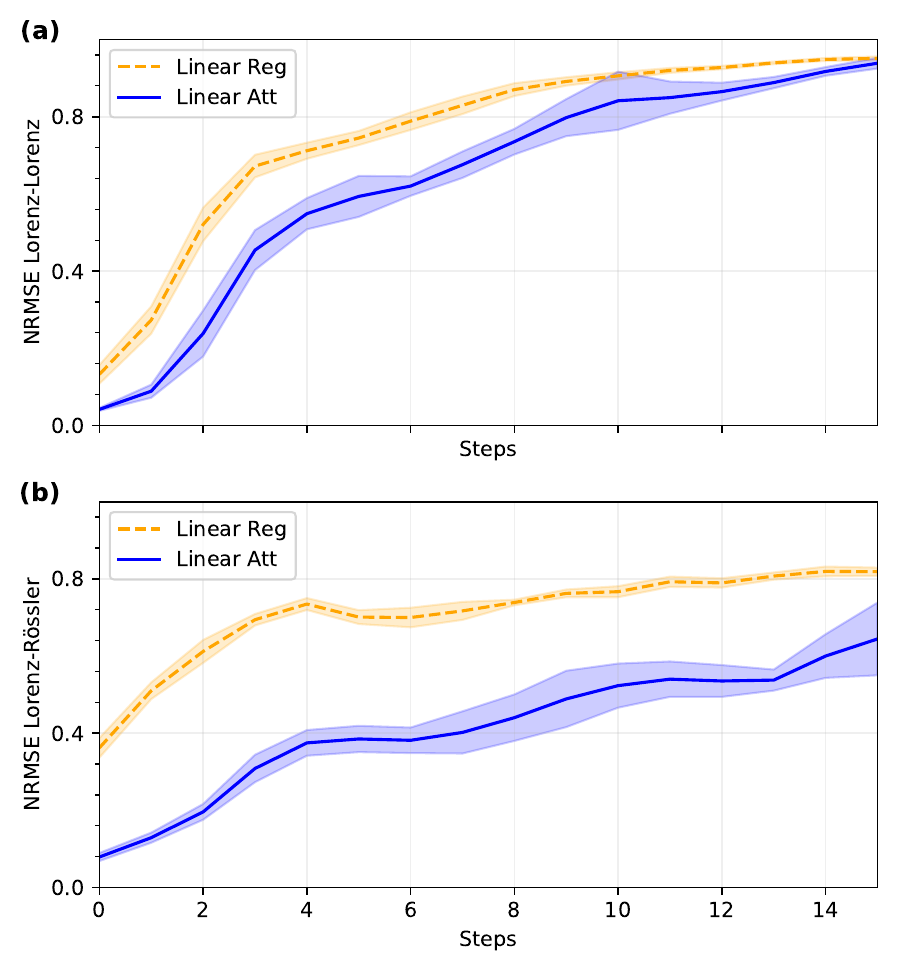}
	\caption[]{NRMSE for (a) UCTLS and (b) ALRS for classical linear regression (dashed orange line) and attention-enhanced reservoir computing (solid blue line) with varying the number of steps for future prediction. The shaded areas depict the standard deviations of NRMSE and VPT over ten time series. The photonic reservoir size is set to $N=50$ nodes.}
	\label{fig:NRMSE_over_steps}
\end{figure}

Finally, we analyze the NRMSE for predicting more than one-step ahead for the UCTLS and ALRS in the open-loop configuration \cite{Takano:18}.
The results for UCTLS are shown in Fig. \ref{fig:NRMSE_over_steps}(a) and for ALRS in Fig. \ref{fig:NRMSE_over_steps}(b), depicting the NRMSE over multi-steps for future predictions.
The UCTLS system is simulated with a driving force of $\sigma_{\text{force}}=0.05$.
Both reservoirs for the UCTLS and ALRS tasks utilize 50 nodes.
The dashed orange line shows the classical ridge regression approach while the solid blue line depicts the attention-enhanced scheme.
The improvement for the UCTLS case is clear for all prediction steps into the future, decreasing the NRMSE by a small amount.
More impressively, the NRMSE decreases for the ALRS in Fig. \ref{fig:NRMSE_over_steps}(b), where a significant reduction in the NRMSE is clearly observed.

\section{Conclusion}
\label{sec:conc}

We introduced an attention mechanism to reservoir computing to demonstrate a notable improvement for predicting complex dynamical systems. Across our simulations of the UCTLS and ALRS, the attention-enhanced reservoir consistently outperformed classical linear regression methods. It has shown particular advantages in scenarios with smaller reservoir sizes, indicating its potential to reduce system requirements without sacrificing performance.
This result is particularly advantageous for integrated photonic reservoirs, where a small reservoir can be easily implemented.
It also highlights the possibility of a full photonic implementation, as a photonic accelerator.

This study shows the importance of dynamically-adjustable weights, characteristics of the attention mechanism, and their effect in predicting the complex behaviors of nonlinear dynamic systems. Notably, the attention-enhanced reservoir exhibits swift adaptability to sudden changes in dynamic input, a quality that promises significant improvements in real-time applications.
By achieving lower prediction errors and capturing key spectral features more accurately, the attention-enhanced approach provides an efficient method for integrating attention within various reservoir computing schemes.

\begin{acknowledgments}
This study was partly supported by JSPS KAKENHI (JP19H00868, JP20K15185, JP22H05195). We thank Ryugo Iwami, André Röhm, and Dhruvit Patel for productive discussions.
\end{acknowledgments}

\appendix

\section{Multidimensional Target Signal}
\label{sec:prediction_inference}

We clarify how the prediction of an $m$-dimensional target signal for the linear attention-enhanced reservoir computer works in detail.
From Eqs. \eqref{eq7} and \eqref{eq8} we get the modified versions for $M$ dimensions as follows:
\begin{align}
    \textbf{w}_{\text{att,l,m}} &= \textbf{W}_{\text{net,m}} \textbf{r}_l, \label{Aeq1} \\
    d_{l,\text{m}} &= \textbf{w}_{\text{att,l,m}}^T \textbf{r}_l, \label{Aeq2}
\end{align}
Here the matrix $\textbf{W}_{\text{net,m}} \in \mathbb{R}^{N \times N}$ consists of the weights that are trained via the gradient descent algorithm, where $N$ is the number of reservoir nodes. The index $m$ runs over all target dimensions $M$.
The matrix $\textbf{W}_{\text{net,m}}$ is multiplied with the standardized reservoir state vector $\textbf{r}_l$ yielding the attention vector for the $m$-th dimension $\textbf{w}_{\text{att,l,m}} \in \mathbb{R}^{N\times1}$.
Thus every single dimension of the target system has its own attention weight vector $\textbf{w}_{\text{att,l,m}} \in \mathbb{R}^{N\times 1}$, and thus its own attention layer weights $W_{\text{net,m}} \in \mathbb{R}^{N \times N}$ matrix.

After computing the attention weights, $\textbf{w}_{\text{att,l,m}}$ is multiplied with the reservoir state $\textbf{r}_l$ to yield the final prediction $d_{l,\text{m}}$.

\section{Nonlinear Attention}
\label{app:non_att}

\begin{figure}%
	\centering
   	\includegraphics[width=0.5\textwidth]{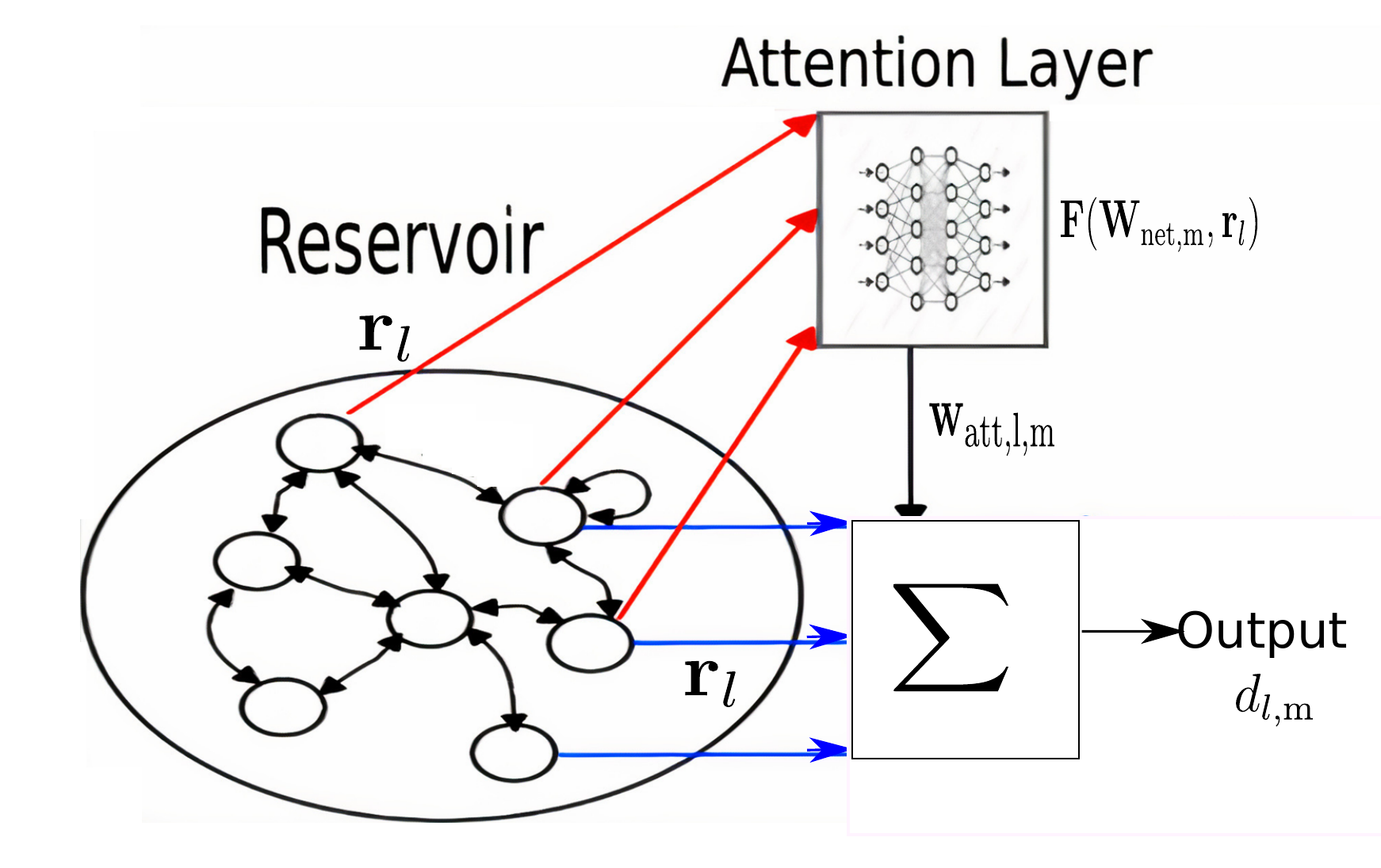}
	\caption[]{Schematic of the attention-enhanced reservoir computing setup with a nonlinear attention layer. The attention layer is replaced by a neural network, that yields the attention weights $\textbf{w}_{\text{att,l,m}}$, which are then weighted with the reservoir node states to generate an output. }
	\label{fig:sketch_nonliner_attention}
\end{figure}%

In our proof-of-concept demonstration, we used a linear transformation as a simplest form of a "neural network" for the attention layer.
To show a more complex architecture, we use a 3-layer feed-forward neural network to compute the attention weights for every dimension of the target system, i.e. $M=3$ for the tasks in this paper.
A sketch of the nonlinear attention approach is shown in Fig. \ref{fig:sketch_nonliner_attention}.
We can alter Eq. \eqref{Aeq1} and Eq. \eqref{Aeq2} accordingly as follows:\\
\begin{align}
    \textbf{w}_{\text{att,l,m}} &= F(\textbf{W}_{\text{net,m}},\textbf{r}_l), \label{eq7nonling} \\
    d_{l,\text{m}} &= \textbf{w}_{\text{att,l,m}}^T \textbf{r}_l, \label{eq8nonling}
\end{align}\\
Here $F(\textbf{W}_{\text{net,m}},\textbf{r}_l)$ is a neural network that takes $\textbf{r}_l$ as input and has $W_{\text{net,m}}$ trainable weight parameters. The input layer consists of the $N$ reservoir states, while the remaining architecture can be chosen freely. 
One hidden layer consisting of $N$ nodes is chosen.
The output layer of each network has $N$ nodes, which compute the $N$ attention weights $\textbf{w}_{\text{att,l,m}}$ for each target dimension $m \in \{M\}$.
Collecting all weights, the attention layer consists of $6N^2$ weights, where $N^2$ comes from the interlayer weights, and the factor $6$ results from $3$ networks all having $2$ interlayer weights.

\begin{figure}[t]%
	\centering
   	\includegraphics[width=0.5\textwidth]{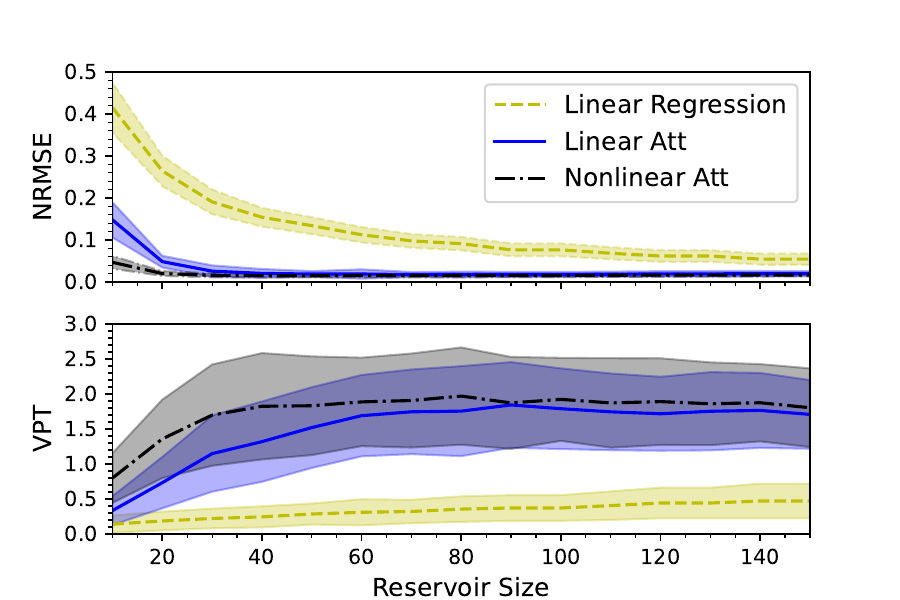}
	\caption[]{(a) NRMSE and (b) VPT for classical linear regression (dashed orange line), linear attention (solid blue line), and nonlinear attention-enhanced reservoir computing (dashed-dotted black line) with varying reservoir size $N$ for UCTLS with a forcing strength of zero, essentially making it a classic single Lorenz task. The shaded areas depict the standard deviations of NRMSE and VPT over ten time series.}
	\label{fig:VPT_NRMSE_nonlinear}
\end{figure}

We use this approach as a nonlinear attention, and compare its performance for our UCTLS system to the linear attention and the ridge regression approach over the reservoir size.
The results are depicted in Fig. \ref{fig:VPT_NRMSE_nonlinear}.
It is shown that the a significant lower NRMSE and higher VPT is achieved, especially for smaller reservoirs.
This advantage is reduced for larger reservoirs, stemming from the limited complexity of the problem.
Therefore, future plans involve to apply the attention-enhanced reservoir to more complex tasks.

\section{Convergence of the Loss}
\label{sec:loss}

The attention-enhanced reservoir computer loses the advantage of the analytically solvable ridge regression approach, because a gradient descent algorithm has to be used.
The output-layer optimization of reservoir computing is not a limiting factor due to the recent advances in hardware acceleration.
Considering the amount of research conducted to develop photonic-based accelerators for neural networks, we see no disadvantage using a gradient descent, in optimizing the output layer.
In addition, the advantage of ridge regression only consists for small reservoirs due to the complexity of ridge regression compared to a gradient-descent-based approach.
The order of the ridge regression is roughly $\mathcal{O}(L^3)$, where $L$ is the number of parameters used. On the contrary, the order of the gradient descent algorithm is $\mathcal{O}(LKT)$, where $K$ is the number of steps needed for convergence, and $T$ is the number of training data points.
For our approach, the two complexities thus become $\mathcal{O}(N^3)$ and $\mathcal{O}(N^2KT)$ for the ridge regression and the gradient descent approach respectively, where $N$ is the size of the reservoir.
Thus, generally for large $L$ and $N$, the gradient descent algorithm may be faster, depending on the number of data points and epochs for convergence.
We believe that an increase in reservoir computer size with an increase in the output layer may enable the reservoir computer to tackle more complex tasks.

\begin{figure}[t]%
	\centering
   	\includegraphics[width=0.5\textwidth]{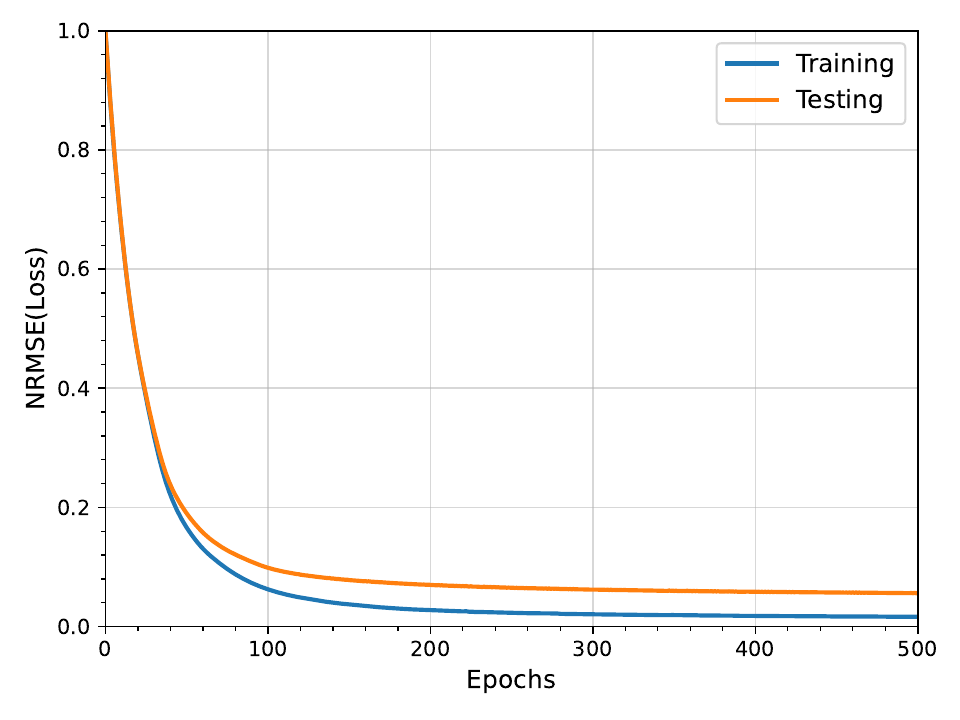}
	\caption[]{Training and testing NRMSE (loss) over the epochs of the gradient descent algorithm.}
	\label{fig:Loss_over_epochs}
\end{figure}

To test the complexity, we plotted the NRMSE over the number of epochs for an 50-node reservoir ($N=50$) using the linear attention approach, as shown in Fig. \ref{fig:Loss_over_epochs}.
It is clear that the NRMSE exponentially and asymptotically converges to a minimum.

\bibliography{references}

\end{document}